# Analogy of space time as un elastic medium – Can we establish a thermal expansion coefficient of space from the cosmological constant $\Lambda$ ?-

Izabel David[1]

[1] Engineer INSA Rennes and mechanical professor, Paris, France
E-mail : d.izabel@aliceadsl.fr

**Abstract**

This paper advances the state of the art by extending the study of the analogy between the fabric of spacetime and elasticity. As no prior work exists about a potential space-time thermal expansion coefficient $\alpha$, we explore the analogy of general relativity with the theory of elasticity by considering the cosmological constant $\Lambda$ as an additional space curvature of the structure of space due to a thermal gradient coming from the cosmic web and the cold vacuum and we propose $\left(\frac{\alpha_S \Delta T}{e}\right)^2 = \left(\frac{1}{R_0}\right)^2 = \Lambda$ with $R_0$ being the curvature radius of the space fabric. It follows from this analogy and from the supposed space model consisting of thin sheets of Planck thickness $l_p$ curved by this thermal gradient $\Delta T$ a possible thermal expansion coefficient of the equivalent elastic medium modeling the space $\alpha_S = \frac{l_p \sqrt{\Lambda}}{\Delta T}$ of the order of $\alpha_{space-QFT}$=1.16 x 10$^{-6}$ K$^{-1}$. As space-time and not only space must be considered in general relativity, this paper also proposes an innovative approach which consists in introducing into the interval ds² of special relativity a temperature effect T: $d_s^2 = (1 \pm \alpha_t T)^2 c^2 dt^2 - (1 \pm \alpha_s T)^2 [dx^2 + dy^2 + dz^2]$ (entropy variations correlated with time laps, based on temperature variations affecting always physically the clocks) based on different thermal expansion coefficients for space and time with for the flow of time t: $\frac{ct}{cn\tau} = \frac{k_B t}{nh} \times \Delta T = \alpha_t \Delta T$. With $T \approx 10^6$ Kelvin, n=1, the associate time interval is $4.8 \times 10^{-17}$s and $\alpha_t = 1.0 \times 10^{-6} K^{-1}$. The consequence of these hypothesis is that dark energy potentially becomes a thermal space-time curvature $\left(\frac{\alpha_f T}{l_p}\right)^2$ with f equal to s or t depending of the temperature, the thermal entropy variation of the universe, the Planck thickness and time, that increases since the Big bang, depending on thermal expansion coefficients for space-time $\alpha_s$ and $\alpha_t$ as a function respectively of $\Lambda$, $\frac{k_B}{h} \times t$, in opposition to space-time curvature gravity due to mass/energy density as described in general relativity.

# 1. Introduction

*1.1 Measured strains of space and analogy of the general relativity theory with the elasticity theory*

Einstein's theory of general relativity is over 100 years old and is now widely verified. Thus, spacetime is according to this theory a deformable elastic physical object. Gravitation is thus a manifestation of the geometric deformation of space-time under the effect of the masses or energy density therein. The manifestations of the deformations of this space-time are now known and measured with great precision. We can quote the apparent position variation of stars placed behind the sun during an eclipse measured by Edington **[1]**, the frame dragging of space-time by angular distortion by the rotation of the earth (experiment prob B, Lense-Thirring and frame dragging effects) **[2]**, the simultaneous lengthening and shortening deformations in each of the arms of Ligo/Virgo type interferometers during the passage of gravitational waves **[3][4]**, gravitational lenses or substantial masses located between a galaxy and our field sighting on earth distorts space to the point of making it appear to us in the shape of a circle (a bit like a candle placed behind the flat circle of a stemmed wine glass appears circular by its transparency reflection), and finally the expansion of the universe where the galaxies are "fixed" in a space which expands in an increasingly accelerated way characterized by Hubble's law. All these manifestations of the deformations of space time have led many physicists like A Sakharov **[5]**, J.L Synge **[6]**, C. B Rayner **[7]**, R. Grot **[8]**, V.V Vasilev and L.V Fedorov **[9] [10]**, J.D Brown **[11]**, T.G. Tenev and M.F. Horstemeyer **[12]**, P.A Millette **[13]** and many others as T. Damour in its conferences and books consider that the theory of general relativity is a kind of theory of the elasticity of a deformable elastic space-time medium. We then speak of "elastic metric" or "elastic theory of gravitation". It is within the framework of this analogy that we place ourselves in this paper.

If by analogy, therefore, space-time is considered as an equivalent 4 dimensionnal deformable elastic medium, where a simplistic 2 dimension image is a heavy ball put on a rubber sheet deformed by this ball, they are two consequences. We will study it in the next two paragraphs.

*1.2 The mechanical characteristics of the equivalent elastic medium in the field of the analogy of the elasticity theory with the General Relativity – review of the state of the art*

This equivalent elastic medium must therefore be characterized with the usual parameters linked to all elastic mediums and to the elasticity theory (Young's modulus Y=E, Poisson's ratio ν, density ρ, etc). Thus various authors have sought to establish an equivalent Young's modulus of the space noted $Y_{space}$. We can quote R.Weiss during his nobel prize speech about gravitationnal waves I quote «*In other words, it takes enormous amounts of energy to distort space. One way to say it is, the stifness (Young's modulus) of space at a distortion frequency of 100 Hz is $10^{20}$ larger than steel*».

T.G Tenev et M.F Horstemeyer **[12]** who propose by considering the space-time made up of thin elastic sheets of the thickness of Planck the following formulation (1) giving $Y_{space} = 4.4 \times 10^{113} N/m^2$:

$$Y_{space} = \frac{6c^7}{2\pi \hbar G^2} = \frac{24}{l_p^2 \kappa} \qquad (1)$$

In this expression, c is the speed of light, G is the gravitational constant, $\hbar$ is the reduced Planck's constant, $l_p$ is Plank's length (thicknesses of the thin sheets supposed to constitute the space fabric in **[12]**) and κ is Einstein's gravitational constant ($\kappa = \frac{8\pi G}{c^4}$).

M. Beau proposes a space bulk modulus **[14]** and arrives at $K_{space} = 1.64 \times 10^{109} N/m^2$.

K. McDonald **[15]** proposes another expression of the Young's modulus (2) based on dimensional equations and obtain:

$$Y_{space} = \frac{c^2 f^2}{G} \qquad (2)$$

Where f is the frequency of the gravitational wave. He thus obtains for a gravitational wave of 100 Hz $Y_{space} = 10^{20}$ Y$_{steel}$ so, $Y_{space} = 4,5 \times 10^{31} Pa$ as R. Weiss.

D.Izabel in **[16]** arrives at an expression similar to K.Mc Donald by studying the elastic deformations in space dynamics located in the arms of the Ligo/Virgo laser interferometers and by studying the elastic deformations of a space cylinder twisted by the rotation of two black holes. He thus obtains the expression (3) of the Young's modulus of space similar to K.McDonald at the factor π close:

$$Y_{space} = \frac{\pi f^2 c^2}{G} \qquad (3)$$

This leads by considering a density ρ from quantum field theory (hypothesis similar to T.G Tenev and M.F Horstemeyer **[12]**) at $Y_{space} = 1$ at $4.0 \times 10^{113} Pa$.

A.C. Melissinos **[17]** considers vibrating plates in the planes of the arms of the interferometers and arrives at the following expression (4) of the Young's modulus of space:

$$Y_{space} < \frac{\pi c^2 f^2}{4G} \times \frac{c \Delta \tau}{\Delta z} \qquad (4)$$

In this expression, Δτ is the length of the Gravitational Wave burst and the total path length traversed by the GW is designated by Δz. The numerical application is done with Δτ ≈ 1 s and Δz ≈ 400 Mpc and leads it to a next value of Young's modulus $Y_{space} < 2.5 \times 10^{-17} (c^2 f^2/G)$.

Finally, let us quote S.R. Hwang in **[18]** which by comparing the energy of gravitational waves for different frequencies (35 to 100 Hz and different GW) with the deformation energy of a spring modeling the lengthening and shortening of the arms of the interferometers arrives at values of the Young's modulus of space time $Y_{space} = 1.0 \times 10^{36} \, at \, 1.0 \times 10^{54}$ Pa.

Another key parameter of any elastic medium is of course the coefficient of transverse deformation, the space Poisson's ratio ν. Again, many authors have proposed values. From the simultaneous deformations of the arms of the Ligo/Virgo interferometers (while one arm is shortened by a deformation by shortening of the order of $-10^{-21}$ the other lengthens by the same amount of $+10^{-21}$) T.G Tenev and M.F Horstemeyer **[12]** propose ν=1 which presupposes a certain anisotropy of space which behaves like a kind of thousand leaves during the passage of a gravitational wave , each plane deforming successively according to the polarizations $A^+$ and $A^\times$. The Poisson's ratio being close to zero in the direction of propagation of the wave and being equal to 1 in the plane perpendicular to the direction of propagation.

D.Izabel in **[16]** arrives at the same conclusion on this Poisson's ratio.

Concerning the equivalent density of the medium space ρ, A Sakharov **[5]** shows that quantum considerations of space (quantum field theory) make it possible to go back to an elastic theory of space (5). We quote it below:

"In Einstein's theory of gravitation, it is postulated that the action of space-time depends on the curvature (R is the invariant of the Ricci tensor):

$$S(R) = -\frac{1}{16\pi G} \int (dx)\sqrt{-g}R \qquad (5)$$

The presence of action (1) results in a "metric elasticity" of space,that is, the generalized forces that oppose the curve of space."

T.G Tenev and M.F Horstemeyer **[12]** and D. Izabel **[16]** also follow this path which leads them by considering the minimum non-zero energy of the vacuum to the following expression (6):

$$\rho = \frac{Y_{space}}{4c^2} = 1.3 \times 10^{96} \frac{kg}{m^3} \qquad (6)$$

P.A Millette in **[13]** formula 19.36 proposes the following expression (7) for the density of space:

$$\bar{\rho}_0 = \frac{32c^5}{\hbar G^2} = 1.7 \times 10^{98} kg/m^3 \qquad (7)$$

Finally concerning the shear modulus of the middle space P.A Millette always in **[13]** formula 19.14 and 19.22 proposes the following expression (8a) :

$$\bar{\mu}_0 = \mu = G = \frac{Y}{2(1+\nu)} = \frac{32c^7}{\hbar G^2} 1.5 \times 10^{115} N/m^2 \qquad (8a)$$

For bulk modulus (formula 19.21 of **[13]**) and 8b below:

$$\bar{\kappa}_0 = K = \frac{\bar{\mu}_0}{32} = \frac{c^7}{\hbar G^2} = \lambda + \frac{2}{3}\mu = \frac{Y}{3(1-2\nu)} = 0.046875 \times 10^{115} N/m^2 \qquad (8b)$$

That concludes this review of the characteristics of space following the analogy of it as an equivalent elastic medium. Two conclusions emerge from this state of the art.

First, the mechanical parameters of the equivalent elastic material constituting in our analogy the comic fabric are not of the same magnitude as those found on Earth given the orders of magnitude extremely small of the strains of ($10^{-21}$), extremely large of the Young's modulus ($10^{113} Pa$), the Poisson's ratio outside the usual standards ($\nu$=1), the density ρ of the medium also extremly large ($10^{98} kg/m^3$) and its associated anisotropy linked to the Poisson's ratio if it is 1 in a direction perpendicular to the propagation of the wave.

Secondly, and this is partly the subject of this paper, no publication to our knowledge deals with a possible expansion coefficient α of this space medium. We are therefore going to determine an original and innovative approach in this paper to try to propose a mechanical expression and a numerical value of this possible expansion coefficient α of the equivalent spatial fabric on the one hand and to study the consequences on time on the other hand.

*1.3 Analogy about the behaviour law of the space following the general relativity with the Hooke's law in elasticity without and with cosmological constant Λ – thermal gradient implication about the different curvatures that have to be considered*

The second aspect about this analogy is the law governing this equivalent elastic medium, this very special space fabric. Indeed, it is well known that Einstein's equation of general relativity without cosmological constant Λ relates the curvature of space $G_{\mu\nu}$ to the density of energy which deforms it $T_{\mu\nu}$ see equation (9).

$$G_{\mu\nu} = R_{\mu\nu} - \frac{1}{2} g_{\mu\nu} R = -\frac{8\pi G}{c^4} T_{\mu\nu} \qquad (9)$$

In this expression $R_{\mu\nu}$ is the Ricci tensor resulting from the contraction of the Riemann tensor, R is the scalar curvature resulting from the contraction of the Ricci tensor and $T_{\mu\nu}$ the momentum energy tensor. μν varying from 0 for time to 3 for the 3 dimensions of space.

D.Izabel showed in **[16]** that the general relativity equation in 4 dimensions presents an analogy with respectively the expressions in one and two dimensions of the curvature of beam (10, 11) and plate (12) in pure bending under two moments applied at each extremity (see Figure 1.) according to the Timoshenko's strength of materials theory issued of the elasticity theory.

$$\frac{1}{R^2} = \frac{2}{EI}\left(\frac{W_{ext(total)}}{L}\right) = K\left(\frac{W_{ext(total)}}{L}\right) \quad (10)$$

In analogy with :

$$G^{\mu\nu} = -\frac{8\pi G}{c^4}(T^{\mu\nu}) = -\kappa(T^{\mu\nu}) \quad (11)$$

In these expressions, R is the radius of curvature of a beam of span L, moment of inertia I and Young's modulus E=Y associated with a work of the external forces W. K is the mechanical coupling constant between the curvature and the strain energy U of the beam which is equal to the work applied external forces W.

Or in theory of the plates of thickness h following **[25]** with $R_x$, $R_y$, $R_{xy}$, the curvature radii of the plate in bending in the different directions, $\Delta x \cdot \Delta y \cdot t$ an elementary volume of the plate of thikness t, $\nu$ the Poisson's ratio and E the Young's modulus of the material constituting the plate.

$$\left[\left(\frac{1}{R_x}\right)^2 + \left(\frac{1}{R_y}\right)^{.2} + 2(1-\nu)\left\{\left(\frac{1}{R_{xy}}\right)^2\right\} + 2\nu\left\{\frac{1}{R_x}\frac{1}{R_y}\right\}\right] = \frac{24(1-\nu^2)}{Et^2} \times \frac{\Delta U}{t\Delta x\Delta y} \quad (12)$$

T.G Tenev et M.F Horstemeyer **[12]** but also T. Damour in his book "if Einstien was told to me ", showed this analogy from a tensor point of view in terms of Hooke's law (13) and (14) with $\sigma^{kl}$ the stress tensor, $\varepsilon^{ij}$ the strain tensor, Y the Young's modulus and $g^{kl}$ a metric.

$$\sigma^{kl} = \frac{Y}{1+\nu}\left(\frac{\nu}{1-2\nu}g^{ij}g^{kl} + g^{ik}g^{il}\right)\varepsilon_{ij} \quad (13)$$

$$T_{\mu\nu} = \frac{1}{\kappa}\left(R_{\mu\nu} - \frac{1}{2}Rg_{\mu\nu}\right) \quad (14)$$

In the space fabric model of T.G Tenev et M.F Horstemeyer **[12]**, space is assumed to be made up of ultra-thin sheets of Planck thickness, having an elastic behavior. It is in this space model that we will place ourselves in the rest of this publication.

But the tensorial equation of Einstein (9) can be also written by considering this time the cosmological constant Λ as a materialization of a certain repulsive dark energy. It is written as follows (see formula (15) and (16)):

$$R_{\mu\nu} - \frac{1}{2}g_{\mu\nu}R = \frac{8\pi G}{c^4}T_{\mu\nu} - \Lambda g_{\mu\nu} \qquad (15)$$

Or by factoring Einstein's constant κ :

$$R_{\mu\nu} - \frac{1}{2}g_{\mu\nu}R = \frac{8\pi G}{c^4}\left[T_{\mu\nu} - \frac{c^4\Lambda}{8\pi G}g_{\mu\nu}\right] \qquad (16)$$

But the Einstein's tensorial equation (9) considering the cosmological constant Λ can be also written as an additional curvature present in all space (frame of this paper). It is written as follows formula (17):

$$G_{\mu\nu} = R_{\mu\nu} - \frac{1}{2}g_{\mu\nu}R + \Lambda g_{\mu\nu} = -\frac{8\pi G}{c^4}T_{\mu\nu} \qquad (17)$$

If we follow completely the analogy of the beam or the plate in elasticity describes before it exists not one but two sources of curvature (see chapter 3): one under the applied masses developped in **[12]** and **[16]** and one under the temperature gradient (so with 0 mass), hence our idea of associating the second with the cosmological constant if our analogy is correct, the first one having already been partially demonstrated in **[16]**.

So, we come to the subject of this publication. In the same way that we have shown that Einstein's constant κ, by analogy with an elastic medium made of thin sheets of thiknesses $l_p$, (plate theory) could be expressed in terms of mechanical constants **[12] [16]** see expression (12) ($\frac{24(1-\nu^2)}{Et^2} \to \kappa \to Y = \frac{24}{l_p^2\kappa}$), the cosmological constant Λ which is generally associated with a dark energy **[19]** opposed to gravitation (15 and 16), can it not also be expressed in terms of mechanical parameters as an additional curvature present in all space (17) due at a thermal gradient applied at these thin sheets of thicknesses $l_p$?

This constant Λ generally associated with an expansion of space **[20] "Dark energy as a curvature of space-time induced by quantum vacuum fluctuations"**, can it not be correlated with a hidden mechanical behaviour of space via a parameter missing from the state of the art cited above in the analogy of space as a elastic medium, namely a thermal expansion coefficient α of it? Does not the phenomenal temperature difference between the cosmic web that fills the entire universe and the icy vacuum at -2.73 K constitute such a thermal gradient sufficient to impose on the sheets **[12]** constituting the space of a thickness of Planck in the framework of our model an additional source curvature of this cosmological constant? It is these hypothesis that we will study in this paper, thus placing ourselves in the continuity of the publications and theoretical models of T.G Tenev and M.F Horstemeyer **[12]** and D. Izabel **[16]**.

## 2. Methods

The following methodology has been implemented to estimate, within the framework of the analogy of space as an equivalent elastic medium, the value of a possible thermal expansion coefficient α of space in line with the cosmology constant Λ associated with an additional thermal space curvature in the framework of a Planck's thickness sheet space model as considered by T.G Tenev and M.F Horstemeyer **[12]**:

1) Within the framework of the analogy of an elastic medium to model the deformations of space in the presence of mass energy, restructure the simplified and appropriate mechanical models among those already developed to evaluate a possible generalized thermal curvature of this one,

2) Search for certain scientific data that can feed this mechanical model of curved space under the effect of a thermal gradient between the cosmic web and the space vacuum,

3) Consider the cosmological constant Λ not as a gravitationally repulsive dark energy (if placed to the right of equations (15) and (16) but as an additional curvature present in all space (if placed to the left of the equation (17)) and analysis of the consequences of this approach,

4) Extract from the mechanical model of curvature of space under thermal gradient and from the scientific data available in connection with this model by considering the cosmological constant Λ, a possible thermal expansion coefficient α of the space medium,

5) Discuss the representativeness of this model by varying the assumptions (values of Λ) and seeing the consequences on the value of the thermal expansion coefficient of the associated elastic space fabric,

6) Discuss the additional verifications necessary to confirm this approach,

7) Investigate by the state of the art analyse, the potential effect of temperature on time via entropy variation and its profound implication on the interval $ds^2$ in special relativity, value of Λ in general relativity expressed in terms of curvature of space and time,

8) Presentation of the possible consequences of this model concerning the interpretation of dark energy,

9) Proposition of a testing way to verify this theory.

## 3. Search for a simplified mechanical model of space allowing to represent a thermal curvature of this one

*3.1 Analogy of the thermal curvature of an elastic medium with a simplified approach of resistance of materials in one dimension - beam under thermal gradient*

In the strength of materials, two phenomena and only 2 can create curvature:

- masses supported by the object,
- a thermal gradient applied to the object (beam, plate, shell).

We studied the first type of curvature in **[16]** and showed that masses placed on a beam create curvature (18). Thus we have shown in **[16]** that the curvature of a beam in pure bending (solicited by two moments M at each end) (18) takes a form similar to Einstein's equation from the point of view of the analogy of space as an elastic medium in the form curvature =K x a linear strain energy density U/L:

$$\frac{1}{R^2} = \frac{2}{EI}\left(\frac{U}{L}\right) \qquad (18)$$

With U the strain energy of the beam, L its span, E the Young's modulus of the material, $I = \frac{be^3}{12}$, its moment of inertia, $(b \times e)$ its section and R its curvature radius (See Figure.1).

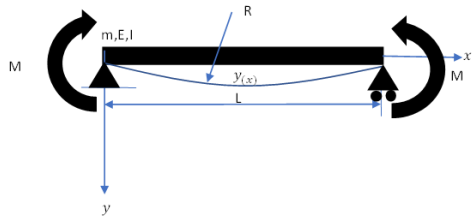

**Figure 1 : Definition of a Timoshenko's beam in pure bending -**

The second type of curvature, we study it in this article, it is the curvature created by the difference in temperature between the two extreme fibers of the beam. The beam is then subjected to a thermal gradient ΔT which leads to an elongation of the heated fibers and a shortening of the cooled fibers. These differences in elongation create a curvature without internal forces of the beam if this one is not constrained in displacement somewhere along its surface.

T. Damour tells us in his various conferences and publications **[9] [11]** that curvature in the Einstein sense has the dimension of an angle divided by a surface (see also the first definition of curvature and differential geometry in C. F Gauss work **[21]**). We will show that is indeed the case of a beam loaded by a thermal gradient (temperature difference T between the lower and upper fibers of the beam $\Delta_T = T_{ext} - T_{int}$) with $T_{ext}$ and $T_{int}$ the temperature of each side of the sheet considered.

Let us now consider the case of an identical beam in pure bending undergoing a thermal gradient (See. figure. 2):

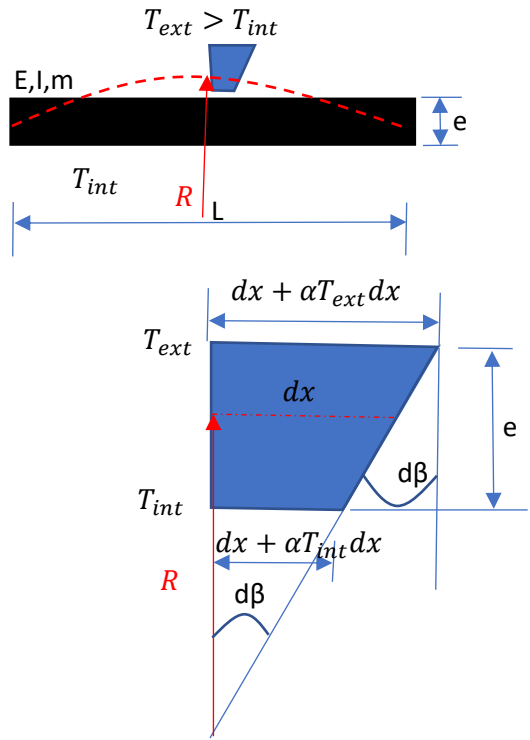

**Figure 2 : Strain of a beam in bending under a thermal gradient ΔT –**

The strain energy U (19) is always with M a bending moment for a beam of span L, thickness e and rigidity EI:

$$U = \frac{1}{2}\int_0^L \frac{M^2}{EI} dx \qquad (19)$$

Figure 2 and formula (20) allow us to write the following geometric relations between the angle β (radiant) and the curvature (1/R) of the beam with h the height of the beam, α its expansion coefficient and ΔT the thermal gradient applied to this beam:

$$\tan(d\beta) = \frac{dx}{R} = \frac{(dx+\alpha T_{ext}dx)-(dx+\alpha T_{int}dx)}{e} = \frac{(\alpha T_{ext}-\alpha T_{int})dx}{e} = \frac{\alpha \Delta T dx}{e} \qquad (20)$$

Given the figure. 2, from the relation between the curvature of a beam under a thermal gradient and the second derivative of its displacement equation $y(x)$, we obtain the expression (21):

$$\frac{d^2 y}{dx^2} = \frac{M}{EI} = \frac{1}{R} = \frac{\tan(d\beta)}{dx} = \frac{d\beta}{dx} = \frac{\alpha \Delta T}{e} \qquad (21)$$

Considering the new expression of the moment M from (21):

$$M = EI \frac{d\beta}{dx} \qquad (22)$$

By replacing the moment M by its expression above (22) in the expression of the strain energy U of the beam (19), we then obtain (23):

$$U = \frac{1}{2} \int_0^L \frac{\left(EI\frac{d\beta}{dx}\right)^2}{EI} dx \qquad (23)$$

So, after simplification (EI =constant allong the beam) we obtain (24):

$$U = \frac{EI}{2} \left(\frac{d\beta}{dx}\right)^2 L \qquad (24)$$

After some mathematical calculations, the final result is the equation (25):

$$\left(\frac{d\beta}{dx}\right)^2 = \frac{2}{EI} \frac{U}{L} \qquad (25)$$

That can be compared with the expression recalled above in the case of the beam in pure bending (18).

We therefore have an angle divided by an area as the definition of the curvature (1/R ) squared of the beam (26).

$$\left(\frac{d\beta}{dx}\right)^2 = \frac{d\beta^2}{dx \times dx} = \left(\frac{1}{R}\right)^2 \qquad (26)$$

This corroborates the expression of T. Damour from the differential geometry of Gauss developed by Riemann **[21]** and given again in the formula (27):

$$Curvature = \frac{\alpha + \beta + \gamma - 180°}{Area} \qquad (27)$$

The strain energy of a beam of span L, rigidity EI=YI for a constant thermal gradient ΔT is given by formula (28):

$$U_{\Delta T} = \frac{EI}{2} \left(\frac{d\beta}{dx}\right)^2 L = \frac{EI}{2} \left(\frac{\alpha \Delta T}{e}\right)^2 L \qquad (28)$$

So :

$$\left(\frac{\alpha \Delta T}{e}\right)^2 = \frac{2}{EI} \frac{U_{\Delta T}}{L} \qquad (29)$$

As our analogy is based on the elasticity theory, we can superimpose the load cases and thus superimpose the case of the beam in pure bending (due to the 2 moments M due at masses at each extremity of the beam see Figure 1.) with the case of the beam under thermal gradient uniform constant ΔT (see Figure 2), we then obtain the formula (30) of the generalized curvature of a beam under load and thermal gradient with $W_{ext(total(M+\Delta T))}$ the total external work of the external forces under moment and thermal gradient:

$$\frac{1}{R^2} + \left(\frac{\alpha \Delta T}{e}\right)^2 = \frac{2}{EI}\left(\frac{U_M}{L}\right) + \frac{2}{EI}\left(\frac{U_{\Delta T}}{L}\right) = \frac{2}{EI}\left(\frac{W_{ext(total(M+\Delta T))}}{L}\right) \qquad (30)$$

This expression is compatible from the point of view of the elastic analogy with the expression (17) of the Einstein's field equation with cosmological constant Λ according to the correspondences (31) and (32) if one assumes ( hypothesis of this article) that the cosmological constant Λ is by analogy correlated with a thermal gradient applied in all space:

- For the masses curvature the analogy between general relativity and strength of material/elasticity is:

$$R_{\mu\nu} - \frac{1}{2}g_{\mu\nu}R \rightarrow \frac{1}{R^2} \qquad (31)$$

- For the thermal curvature the analogy between general relativity and strength of material/elasticity is:

$$\Lambda g_{\mu\nu} \rightarrow \left(\frac{\alpha \Delta T}{e}\right)^2 \qquad (32)$$

About the formula (31), indeed, for memory, it can be prove that the Ricci tensor for a classical 2sphere is 2/R² (see book what is space time made of? of D.Izabel).

*3.2 Analogy of space as an elastic medium with the thermal curvature of a thin plate associated according to Timoshenko's theory*

Considering space as a fabric made up of thin sheets has already been explored as we have said above by many authors. Let us quote Melissinos **[17]**, T.G Tenev and M.F Horstemeyer **[12]** D. Izabel **[16]**, H A Perko **[22]**.

It is therefore quite natural that we take up this hypothesis on the structure of the fabric of space.

Moreover, if we take a piece of the universe locally, its surface will be considered almost flat according to the value of the cosmological curvature k=0 obtained by measurements from the Planck satellite **[23]**. In these publications, it is proven that the joint constraint with BAO measurements on space curvature is consistent with a flat universe,

Ωk$=-kc^2(r_0 H_0)^{-2}$ = -0.0010 +0.0018 / -0.0019."

As a reminder, the space expansion scale factor is written $R_{(t)} = a_{(t)} = \frac{r_{(t)}}{r_0}$. $r_0$ is the radius of reference and $r_{(t)}$ the radius of the 3sphere in the metric of Friedmann-Lemaitre-Robertson Walker.

The Hubble constant squared (1/s²) is : $H_0^2 = \frac{8\pi G}{3}\rho_c$ with $\rho_c$ the critical density of the medium.

So, the curvature k (unit 1/m²) of the univers can be considered as flat (33).

$$\Omega_k = -kc^2(r_0 H_0)^{-2} = -\frac{3kc^2}{8\pi G \rho_c a^2} \qquad (33)$$

$\Omega_k$ is therefore the dimensionless curvature parameter of space (34):

$$\frac{\frac{1}{m^2} \times \frac{m^2}{s^2}}{\frac{m^3}{kg s^2} \times \frac{kg}{m^3} \times 1} = 1 \qquad (34)$$

It is therefore not absurd to consider space as an infinitely long thin plates superposition given the very large radius of curvature of the universe associated with its gigantic size **[24]**.

In this case the curvature of a thin plate under a thermal gradient is well known and is given in S. Timoshenko in his book **[25]** in chapter 14 formula 50. We give the expression below (formula 35) similar to that obtained in the case of a beam :

$$\frac{\alpha \Delta T}{e} = \frac{1}{R} \qquad (35.a)$$

Or squared to stay consistent with the previous paragraph regarding beam theory (35.b):

$$\left(\frac{\alpha \Delta T}{e}\right)^2 = \left(\frac{1}{R}\right)^2 \qquad (35.b)$$

In these two expressions the curvature squared of the plate $\left(\frac{1}{R}\right)^2$ is therefore linked to the thickness of the considered plate e, to the thermal expansion coefficient α associated with the elastic material and to the thermal temperature gradient ΔT between the extreme fibers of the plate.

It is therefore this model that we will consider later in the continuity of the authors cited above **[12]**, **[16]**. We must therefore establish the different parameters involved in this model. Namely, what thickness e of the sheets? what intensity of the thermal gradient ΔT?, what value of the curvature 1/R?. This is what we will study in the next chapter.

**4. Search for certain scientific data that can feed this mechanical model of thermal curvature of the space fabric**

*4.1 What plate thickness consider?*

We know from A. Sakharov **[5]** that the potential quantum nature of vacuum can generate an elastic metric of space. T.G Tenev and M.F Horstemeyer in **[12]** consider Planck thickness sheets. D. Izabel in **[16]** manages to find the Young moduli of the different authors by considering also a Planck length. P.A Millette **[13]** does the same.

Consequently, we will consider as in **[12]** within the framework of this publication, a thickness e of plate or fiber of the elastic fabric of space equal to $l_p$ the length of Planck (36).

$$e = e_p = l_p = \sqrt{\frac{\hbar G}{c^3}} \qquad (36)$$

**Remark**

This model of elastic fabric of thickness $l_p$ have been studied in the thesis of T. Tenev to modelize the sun gravity. There is a perfect accordance between theory and model **[12]** and **[30]**.

*4.2 Which thermal gradient consider?*

New data show a temperature gradient between the absolute vacuum at 2.73 K and the cosmic web see **[26]** "The Cosmic Thermal History Probed by Sunyaev–Zeldovich Effect Tomography". In this article we can read:

*"We estimate Te, the density-weighted electron temperature of the universe, which goes from $7 \times 10^5$ K at z=1 to $2 \times 10^6$ K today" The cosmic thermal history probed by Sunyaev-Zeldovich effect tomography YI-KUAN CHIANG , RYU MAKIYA, BRICE MÉNARD ET EIICHIRO KOMATSU " [26].*

In this paper, the authors therefore highlight a certain thermal gradient between the cold zones of the universe and the very hot zones at the level of the cosmic web. So, this recent article suggests that the average temperature of the gas present in the large structures of the observable Universe has been multiplied by 10 during the last ten billion years to reach about two million Kelvin today.

We will therefore retain this hypothesis*: ΔT=2,000,000 °K.*

*4.3 What curvature of space associated with this thermal gradient consider?*

P. J. E. Peebles in his paper **[19]** reviews the different approaches for linking the cosmological constant and dark energy.

Emilio Santos in **[20]** his paper studies the dark energy induced by the curvature of space-time by quantum vacuum fluctuations.

On the basis of these two articles in particular, we will therefore postulate that the cosmological constant is the source of a curvature of space linked to a thermal gradient acting on the thin sheets of space of quantum thickness equal to the length of Planck.

To be consistent with the approaches of A. Sakharov **[5]** and T.G Tenev and M.F Horstemeyer **[12]**, we consider in this study the cosmological constant resulting from vacuum fluctuations (quantum field theory). We will see in chapter 6 how to explore all the scientific options, what gives other values of the cosmological constant Λ- resulting from cosmological observations.

We make this hypothesis based on quantum field theory taking into account all of the previous paragraphs:

1) the analogy of the space fabric as an elastic medium works well with the general relativity equation for mass/energy **[5]** to **[17]**,

2) the elastic analogy suggests two possible curvatures and only two: one resulting from the masses/energy acting in the medium and one resulting from the thermal gradient acting on the medium **[16]**,

3) there is obviously a thermal gradient present in space following **[26]**,

4) Many authors **[12]**,**[16]**,**[22]** and measurements of gravitational waves **[3] [4]** suggest the presence of plane deformations of elastic media (spatial fabric),

5) A.Sakharov **[5]** showed that general relativity is correlated with an elastic metric resulting from quantum fluctuation of space.

*4.4 Final model retained to estimate a coefficient of thermal expansion of the fabric of the equivalent space*

Considering the previous paragraphs, we therefore arrive at the simplified model given in Figure 3 below.

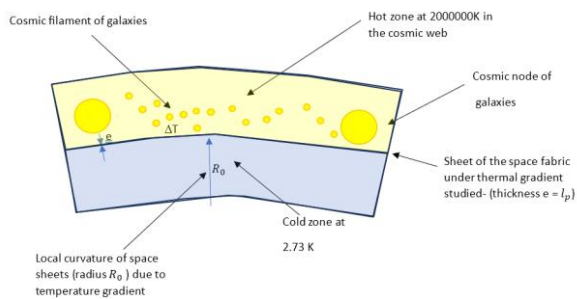

**Figure 3: Simplified mechanical model of elastic space undergoing a curvature by a thermal gradient between a very hot zone and a very cold zone of the universe–**

Thus, the almost flat thin sheets of space (k=0, T of space curvature tending towards infinity **[12]**,**[23]** and **[24]**) of Planck's thickness ($l_p$) according to the reasoning by A. Sakharov **[5]** and T.G Tenev and M.F Horstemeyer **[12]** undergo a thermal gradient resulting from the differences in average temperature between the cosmic web and the space vacuum, **[26]**. This causes their curvature. It is this curvature that the cosmological constant Λ **[19] [20]** can represent. By placing ourselves in the analogy of space functioning as a medium, an elastic fabric, the mechanics of continuous media by the Timoshenko's plate theory **[25]** allows us to model its behavior in a simplified way and to extract a thermal expansion coefficient α of the space fabric.

**Remark**

The thermal conductivity $\lambda$ of the vacuum is by definition 0 W/m².K. This allow us to have a thermal gradient between the two faces of thickness e = $l_p$.

## 5. Consequence of considering the cosmological constant as a generalized thermal curvature

### 5.1 Determination of the space fabric expansion coefficient

From the Timoshenko's expression (36) (see also **[5]**) of a spatial plate curvature (k=0) see **[23]** and **[24]** in pure bending loaded by a thermal gradient ΔT **[26]** and in considering the cosmological constant Λ as an additional curvature (and not a dark energy) associated with the thermal gradient acting in space **[19] [20]** we postulate equation (37):

$$\left(\frac{\alpha_S \Delta T}{e}\right)^2 = \left(\frac{1}{R_0}\right)^2 = \Lambda \qquad (37)$$

$R_0$ is the curvature radius of one of the thin sheet supposed constitute the space fabric **[12]** (see Figure 3).

We deduce from the above expression the thermal expansion coefficient $\alpha_S$ of the spatial fabric:

$$\alpha_S = \frac{e\sqrt{\Lambda}}{\Delta T} \qquad (38)$$

By definition of the comological constant with $\rho_{vacuum}$ the vacuum density and c the speed of light we obtain (39):

$$\frac{c^4 \Lambda}{8\pi G} = \rho_{vacuum} c^2 \qquad (39)$$

So, after some mathematic calculations (40):

$$\Lambda = \frac{8\pi G \rho_{vacuum}}{c^2} \qquad (40)$$

By transferring this expression to the formula of the thermal expansion coefficient of the space fabric (38), we obtain the expression for the space thermal expansion coefficient $\alpha_S$ from the density of the vacuum (41):

$$\alpha_S = \frac{e\sqrt{\frac{8\pi G \rho_{vacuum}}{c^2}}}{\Delta T} = \frac{e\sqrt{\Lambda}}{\Delta T} \qquad (41)$$

If we consider a space plate thickness e of Planck dimension (35) like T. Tenev and M.H Horstemeyer **[12]** and the associated vacuum energy, we obtain (42) a first approximation of this thermal expansion coefficient $\alpha_S$.

$$\alpha_S = \frac{\sqrt{\frac{hG}{2\pi c^3}}\sqrt{\frac{8\pi G \rho_{vacuum,QFT}}{c^2}}}{\Delta T} \qquad (42)$$

So, all calculations performed give (43):

$$\alpha_S = \frac{2G\sqrt{\frac{h\rho_{vacuum,QFT}}{c^5}}}{\Delta T} \qquad (43)$$

With $\hbar$ the reduced Planck's constant ($h/2\pi$), ρ the quantum vacuum density according to quantum field theory, G the gravitational constant, c the speed of light and ΔT the thermal gradient applied to these sheets of space.

Consider the expression of the thermal expansion coefficient $\alpha_S$:

We check that the dimensional equation (44) is satisfied:

$$\alpha_S = \frac{\frac{m^3}{kgs^2}\sqrt{\frac{\frac{kgm^2}{s}\times\frac{kg}{m^3}}{\frac{m^5}{s^5}}}}{K} = K^{-1} \qquad (44)$$

Numerical application using the constants of physics and using the same assumptions as P.A Millette **[13]** and T.G Tenev and M.F Horstemeyer **[12]** gives:

G = 6.6743015 × 10$^{-11}$ m³/kg.s²

h = 6.62607004 × 10$^{-34}$ m².kg/s

$\rho_{vacuum,QFT}$ = 1.11×10$^{96}$ kg/m³

$\Delta T = 2000000\ K$

c = 299,792,458 m/s

We obtain for the thermal expansion coefficient of the space fabric:

$$\alpha_{S,QFT} = 1.16317 \times 10^{-6} \text{ K}^{-1}$$

For memory, the expansion coefficient of steel is worth $12.0 \times 10^{-6}$ K$^{-1}$.

We therefore obtain a result that seems realistic since the space is rather rigid if we refer to the value of κ which represents the flexibility (1/rigidity) of the space and which is equal to $2.0766 \times 10^{-43}$ N$^{-1}$.

We find this result directly from equation (41) with the following data by considering Λ from quantum field theory:

$\Lambda_{QFT} = 2.0717 \times 10^{70}$ m$^{-2}$

$e = l_p = 1.61626 \times 10^{-35}$ m

$\Delta T = 2000000 \text{ } K$

## 6. Discussion of the representativeness of the mechanical model of thermal curvature of space

It is well known that one of the greatest challenges of physics today is this problem of the cosmological constant Λ which, depending on the hypothesis adopted to establish it (vacuum energy resulting from quantum fluctuations in the ground state provided by quantum field theory or cosmological observations via the Planck satellite in particular) leads to a ratio of $10^{120}$ between the two values of Λ! We do not claim in this paper to solve this problem, simply from a scientific point of view it is important to explore what these two values of the cosmological constant imply on the coefficient of potential thermal expansion of the fabric of space.

In this case, the application of formula (41) with the following numerical values:

$l_p = e = 1.61626 \times 10^{-35}$ m

$\Lambda = 1.088 \times 10^{-52}$ m$^{-2}$

$\Delta T = 2000000 \text{ } K$

Leads to an extremely very small coefficient of thermal expansion…

$\alpha_{S,astrophysics} = 8.42936 \times 10^{-68}$ K$^{-1}$

**7. Discussion of the additional verifications necessary to validate or not this model**

To really validate our model, it would of course be necessary to solve this problem of the possible variation of values of the cosmological constant Λ. A model of the universe reproducing the construction of the cosmic web from general relativity and observations exists, however these models do not integrate any mechanical behavior such as the analogy of space as a deformable elastic medium.

It would undoubtedly be necessary to model the space, containing and contained in a sheet structure, to set up the cosmic web and impose the thermal gradient between the hot and cold zones to check whether or not a curvature of thermal origin appears on average. on the entire universe, find out what its intensity is and therefore take advantage of it to evaluate the true value of the cosmological constant and therefore the true value of the thermal expansion coefficient of the space fabric.

**8. Consequence of a thermal expansion coefficient on time for special relativity, general relativity and quantum gravity**

*8.1 Necessary transition from space to space-time in the consideration of temperature*

We have seen in the previous chapters that the analogy with an elastic continuous medium proposes a kind of curvature of zero mass of space related to a temperature effect. But since general relativity is aimed at the curvature of space-time, and not only space, it is necessary to specify how this temperature effect could act on time.

*8.2 Effect of temperature on time*

8.2.1 Generalities on clocks and their physical sensitivities to temperature

It is well known that temperature variations can affect clocks in different ways, especially in mechanical and electronic clocks.

In mechanical clocks, metal parts can expand or contract depending on temperature. This can result in minimal variations in component dimensions, which in turn can affect the accuracy of movement. The materials used in the mechanisms may react differently to temperature changes, which can cause shifts in the clock rhythm. So we have a spatial effect.

In electronic clocks, temperature can influence the frequency of quartz oscillators used to measure time. Quartz crystals have an electromechanical resonance that is sensitive to temperature. Temperature variations can slightly alter the frequency at which the quartz oscillates, which can lead to deviations in the time count. If the equivalent of quartz is very small, see the case of atomic clocks where it is the frequency of an atom that intervenes, we have a quantum effect.

When it comes to electronic processors, temperature variations can also impact their performance. When a processor heats up, its components can expand slightly, which could potentially affect the speed of electrical signals through the circuit. However, modern processor designs usually incorporate temperature control mechanisms to minimize this effect.

So, by thinking about it, the temperature actually influences time by playing on dimensional variations of space or on the vibrations of quartz, so more deeply at the quantum level via $E = h\nu$.

So, to measure time we need clocks which, whatever the physics and the technology used, will be systematically sensitive to the temperature of the environment in which they are immersed. This is the conclusion of this paragraph.

The following question then arises: while it is clear that temperature affects time measuring devices, does it also fundamentally affect time, which is an abstract entity in physics independent from any measuring instruments?

We will see in the following chapter by reviewing the state of the art on this question, how different authors have approached this question by focusing on time as a change in entropy, that is to say at a increased disorder, which in turn is linked to a temperature effect.

8.2.2 State of the art on the link between time, a clock, the entropy variation and the temperature in a black body

In the publications **[27]** and **[29]**, the authors study what is a clock? what kind of measurements can be done? how temperature can affect the time fundamentaly and they explain how time passes according to the increase in entropy and therefore temperature.

In these papers, it is clear that the time itself cannot be defined without a physical process to measure it. I quote **[27]**:

"Einstein pointed out that the definition of time must be based on the clock measure, but it has been pointed out the need of physical meaning for time coordinates . In the analysis of clocks, a timeclock relation has been introduced; it states that there is a conceptually necessary relation between time and a physical process which functions as the core of a clock. This time-clock relation implies that a physical process must exist as the basis of a clock. Time and the physical process cannot be defined independently. The time-clock relation involves also a reference to a physical process in conformity with physical laws. Consequently, a well-defined use of time requires that time has aphysical basis."

So, from this reflection, time is inseparable from a physical process and as in paragraph 8.2.1 we recalled that all physical processes are subject to temperature effects, so time should also be affected by temperature in addition to what Einstein showed in special relativity, namely that time depends on the speed of the observer.

Therefore, the authors in **[27]** formula (1) in **[29]** introduce the definition of the time interval in relation to the local entropy S and the local entropy rate $\dot{S}$, as follows (45):

$$t = \frac{S}{\dot{S}} \tag{45}$$

And from the entropy variation the authors obtain in **[27]** formula (12) and **[29]** a variation τ fonction of the frequency ν of the physical object used to construct a physical clock as required by A .Einstein. However this time this clock depends on the temperature T, the Planck constant h and the Botzman constant $k_B$. They obtain then the fundamental expression (46):

$$\tau_{(T)} = \frac{1}{\nu} = \frac{h}{k_B} \times \frac{1}{T} = \frac{4.799243 \times 10^{-11} [K.s]}{T[K]} \tag{46}$$

An finally according to **[27]** formula (13) and **[29]** in the field of a black body, there is a relation (47) giving the variation of the flow of time t according to the temperature T, the periodic time of oscillation τ and n a natural number.

$$t_{(T)} = n\tau = n\frac{h}{k_B} \times \frac{1}{T} = n.\frac{4.799243 \times 10^{-11} [K.s]}{T} \approx n\frac{4.80 \times 10^{-11} [Ks]}{T} \tag{47}$$

**Remark 1**

In the paper **[27]** table 1 we can see that the effect of the temperature on time is very small.

For T = $10^6$ Kelvin, n=1 the time variation interval linked to this temperature is $4.8 \times 10^{-17}$ s

**Remark 2**

If we take the inverse of the expression (47) and multiply by the time t each side of the equation we obtain (48) that allows to define a time dependant thermal expansion coefficient:

$$\frac{t}{n\tau} = \frac{k_B t}{nh} \times T = \alpha_t T \tag{48}$$

Thus, the term $\frac{k_B t}{nh}$ has the dimension of an expansion coefficient (K$^{-1}$) depending on time which will note $\alpha_t$:

So, the time should be influenced at a quantum level function of the temperature T, the Planck constant h and the Boltzmann constant $k_B$ and thus by the entropy of the medium S. The time is quantified ($n \times h$).

Thus, associated with T=1×10$^6$ Kelvin (see order of magnitude of the temperature of the cosmic web **[26]**), n=1 and considering a time interval of $4.8 \times 10^{-17}$s (following **[27]** table 1), we obtain with (48) an evaluation of the amplitude of this time-dependent thermal expansion coefficient:

$$\alpha_t = 1.0 \times 10^{-6} K^{-1}$$

So, an order of magnitude for $\alpha_t$ similar to that which we obtained for $\alpha_s$ in the first part of our article based on the comological constant $\Lambda$ as a value of the thermal curvature of the fabric of space.

**Remark 3**

In the Cosmic fabric model **[12]**, T.Tenev and M.F Horstemeyer related time lapse to the speed of signal propagation within the fabric.

Thus the variation of the time lapse (49) is written as follows:

$$\frac{d\tau}{dt} = \frac{1}{(1+\varepsilon^{3D})} \tag{49}$$

With the formula (50):

$$\varepsilon^{3D}_{,kk} = c^2 \kappa \rho \tag{50}$$

Where $\varepsilon^{3D}_{,kk} \equiv \nabla^2 \varepsilon^{3D}$ is the Laplacian of the volumetric strain, c is the speed of light, κ is the Einstein constant, and ρ is the density of matter-energy.

In this expression $\varepsilon^{3D}$ is a scalar field that represents the fractional increase of the fabric's mid-hypersurface volume (51):
$$\varepsilon^{3D} \equiv \varepsilon_i^i \tag{51}$$

And the habitual strain tensor is (52):

$$\varepsilon_{ij} = \frac{1}{2}(g_{ij} - \delta_{ij}) \tag{52}$$

Thus the time depend of a variation of volume due to the stress energy tensor that generate strain on the elastic medium (53):

$$\frac{dV}{d\bar{V}} = (1 + \varepsilon^{3D}) \tag{53}$$

But to come back at the complete analogy with the elastic medium this varaition of volume can also come from temperature (54):

$$\frac{dV}{d\bar{V}} = \alpha_S \Delta T \tag{54}$$

And finally in equation (48) we can transpose this to the time as a ratio of length of the space fabric (we see thus the importance of the speed of light as a intrisic caracteristic of the space fabric) see (55):

$$\frac{ct}{cn\tau} = \frac{k_B t}{nh} \times \Delta T \tag{55}$$

8.2.3 State of the art about cosmological time, coordinate time linked to entropy and the dynamics of the expansion of the universe

In **[31]** "Cosmological Time, Entropy and Infinity" the authors go in the same direction as **[27]** and **[29]** by linking time to entropy but go further by associating this variation of entropy with the dynamic temporal evolution of the universe in its entirety from the big bang until now.

I quote "Time is a parameter playing a central role in our most fundamental modelling of natural laws. Relativity theory shows that the comparison of times measured by different clocks depends on their relative motion and on the strength of the gravitational field in which they are embedded. In standard cosmology, the time parameter is the one measured by fundamental clocks (i.e., clocks at rest with respect to the expanding space). This proper time is assumed to flow at a constant rate throughout the whole history of the universe. We make the alternative hypothesis that the rate at which the cosmological time flows depends on the dynamical state of the universe. In thermodynamics, the arrow of time is strongly related to the second law, which states that the entropy of an isolated system will always increase with time or, at best, stay constant. Hence, we assume that the time measured by fundamental clocks is proportional to the entropy of the region of the universe that is causally connected to them. Under that simple assumption, we find it possible to build toy cosmological models that present an acceleration of their expansion without any need for dark energy while being spatially closed and finite, avoiding the need to deal with infinite values."

In this article, the following hypothesis is made, I quote again" the cosmological time *t* measured by such observers is proportional to the entropy of the region of the universe that is causally connected to them."

The autors propose so the following expression of the time function of the entropy at the univers level:

- First the link between the universe entropy and the space temperature T:

Based on the CMB photon gas which the authors **[32]** assume is very close to thermodynamic equilibrium, the entropy S is written as follows (see (56)):

$$S = \frac{4\pi^2 k_B^4}{45 c^3 \hbar^3} V T^3 \tag{56}$$

With V the volume considered (eg horizon $V_{horiz}$), $\hbar$ the reduced Plank's constant, c the speed of light, $k_B$ the Boltzmann's constant and T the temperature associed.

- Secondly the link between the temporal variation of horizon entropy $S_{horiz}$ (57) and univers dynamic expansion:

$$\frac{dS_{horiz}}{dt} = \frac{64\pi^3 k_B^4}{45 \hbar^3} T_0^3 \frac{1}{H_{0,t}^2 \Omega_{0,t}^2} \left( \Omega_{0,t} + R(1 - \Omega_{0,t}) \right) \tag{57}$$

With $T_0$ the present CMB temperature ($T \times R = T_0 \times R_0$ ), R the classical scale factor (sometime noted a also), k the curvature, the Robertson walker metric, t the classical time variable, $H_{0,t}$ (58) the Hubble constant, $\Omega_{0,t}$ (59) the matter density paremeter.

$$H_{0,t} = \frac{1}{R_0} \frac{dR}{dt} \Big|_{t=t_0} \tag{58}$$

$$\Omega_{0,t} = \frac{8\pi G \rho_0}{3 H_{0,t}^2} \tag{59}$$

- Thirdly the link between the entropy variation and the cosmic time $d\tau$ (see 60):

$$\frac{d\tau}{dt} = \frac{dS_{horiz}/dt}{dS_{horiz}/dt \big|_{BB}} = 1 + R\left(\frac{1}{\Omega_{0,t}} - 1\right) \tag{60}$$

With $dS_{horiz}/dt \big|_{BB}$ for the temporal variation of the entropy in the causally connected volume at the BigBang (the coordinate time t (s) that flow at constant rate is equal to τ the cosmological time at the Big Bang of which the unit is the varying length).

- Fourth, the curvature of space which is linked to the interaction of variation of entropy, itself linked to time and temperature:

If $\frac{d\tau}{dt} = 1$ for a flat universe ($\Omega_{0,t} = 1$) the cosmological time flow at a constant rate. It is not the case in the curved space we have (61).

$$\Omega_{0,\tau} = \frac{1}{\Omega_{0,t}} \tag{61}$$

And the curvature k is for $R_0 = 1$ given in (62):

$$k = \frac{H_{0,t}^2(\Omega_{0,t}-1)}{c^2} \tag{62}$$

In conclusion of this state of the art, we find with **[31]** and **[32]**, equation by equation, a direct link between temperature, the evolution of entropy S, the evolution of time t and the associated curvatures k of the universe. We will focus in the following paragraphs on this direct link between temperature and time via an equivalent thermal expansion coefficient of time to be inline with our mechanical analogy of the space-time. Of course, based on **[27] [29] [31]** and **[32]** this link between time and temperature passes in general relativity by the notion of space-time associated this time with curvature of time but also of associated space therefore to a temperature effect. This therefore goes somewhere in the direction of a confirmation of our initial idea of a cosmological constant linked to the thermal curvature of space.

*8.3 Generalization of the special relativity interval taking into account temperature*

Basing of the precendent paragraphs and formula (48) or (55), this invite us to go far and to postulate a rewriting of the interval $ds^2$ of special relativity to introduce a mechanical thermal effect which gives equation (63) with $\alpha_t$ a potentiel dilatation coefficient of the time and $\alpha_s$ a potential dilatation coefficient of space supposed homogenous:

$$d_s^2 = c^2(dt \pm \alpha_t T dt)^2 - (dx \pm \alpha_s T dx)^2 - (dy \pm \alpha_s T dy)^2 - (dz \pm \alpha_s T dz)^2 \tag{63}$$

We postulate that:

$\alpha_s$ is connected with the cosmological constant $\Lambda$ of space by the expression $\alpha_s = \frac{l_p \sqrt{\Lambda}}{\Delta T}$.

$\alpha_t$ is connected with the quantum mecanic and thermodynamic and time following **[27]** and **[29]** by the expression $\alpha_t \approx \frac{k_B t}{nh}$.

Indeed, given the special character and the unknown nature of time, there is no reason in first approach for a coefficient of time dilation to be identical to that of space.

Thus the Expression (63) becomes expression (64) if we separate the spatial part of the time part let :

$$d_s^2 = (1 \pm \alpha_t T)^2 c^2 dt^2 - (1 \pm \alpha_s T)^2 [dx^2 + dy^2 + dz^2] \tag{64}$$

Some important points arises from this expression.

a) The two quantity in the equation (64) above have an opposite sign so a potential contradictory effect,

b) According to the big bang theory, space and time are created at the same time. According to general relativity we arrive at a sungularity at the beginning of the universe where the energy (and therefore the associated mass) is infinitely large and therefore the curvature is infinitely large in an infinitely small volume which requires the introduction of quantum mechanics and therefore the creation of a theory of quantum gravity. But we also often forget, at this moment in the universe the temperature is also infinite and therefore constitutes from the beginning a variable of the problem potentially influencing space-time, therefore space and time in our fabric model cosmic are influenced by temperature. So, it's logical for us to introduce it in the interval formula,

c) Of course, if there is no thermal effect, the interval reverts to the classical interval in spatial relativity.

*8.4 Can curvature, time and temperature be related in general relativity?*

The theory of general relativity in its current version does not directly take into account a slowing effect of time due to heat or temperature, nor does it specifically propose a "thermal curvature of time". This is what we deduce from our previous reflections in Chapter 8.3 based on **[29]** to **[32]**. Indeed, from (64) we have rather a simple dilatation effect and not a curvature effect in potentially all the directions.

However, under certain circumstances, heat can indirectly have an effect on time by altering the curvature of space-time. For example, in very massive and hot astrophysical objects, such as neutron stars or black holes, the effects of heat and pressure can influence the curvature of space-time and therefore the trajectories of moving objects but also the time (time and space become as reversed) (see **[28]**)**.** The Hawking's temperature of black hole, that is the first quantum gravity equation depending on c, h, $k_B$ and G and M the black hole Mass, goes in this direction.

There are also related concepts in theoretical physics, such as string theory and quantum gravity, that attempt to unify general relativity with quantum mechanics. In these theories, it is possible that temperature-related phenomena could have implications for the nature of space-time, but this remains an ongoing area of research and exact specifics are not yet well established.

Thus, based on (64) for the new definition of the interval, taking into account **[27]** "Time & clocks: A thermodynamic approach" and **[29]** "Time and Thermodynamics Extended Discussion on \Time & clocks : A thermodynamic approach" for the temporal variation as a function of temperature and the associated temporal expansion coefficient $\alpha_t$, from equation (37) for the relationship between the cosmological constant Λ and the space expansion coefficient $\alpha_s$, from **[12]** for the thin sheet structure of thickness $l_p$, from **[31]** and **[32]** for the relationship between the temperature, the temporal variation, the entropic variation linked to the expansion of the universe and the associated curvature, we therefore have the following modification of Einstein's equation (17) expressed in terms of mechanical curvature:

$$R_{\mu\nu} - \frac{1}{2} g_{\mu\nu} R - \left(\frac{\alpha_f T}{l_p}\right)^2 g_{\mu\nu} = \frac{8\pi G}{c^4} T_{\mu\nu} \qquad (65)$$

Thus the cosmological constant Λ introduced by Einstein, considered as potentially the source of dark energy, becomes in our article an additional thermal curvature $\left(\alpha_f T / l_p\right)^2$, connected for space and time to two complementary mechanical parameters linked to the space-time fabric, namely its coefficient of thermal expansion of space (f=s) → $\alpha_s$ and its coefficient of thermal expansion of time (f=t) → $\alpha_t$ as well as the sandwich texture of it (multilayer of thickness $l_p$) all depending on the temperature T(→ΔT) acting in connection with the variation of entropy during the temporal evolution of the universe **[31]**.

**Remark 1**

We are not certain today that the dark energy translated into our elastic model in the form of thermal curvature must be strictly constant or not.

Thus, in Einstein's equation the cosmological constant is constant, but in our approach it can potentially vary depending on the temperature or the temperature gradient.

**Remark 2**

The thermal expansion coefficient of time depends on t if we start from the approaches of **[27]** and **[29]**.

*8.5 Synthesis on spatio-temporal approach to consider the temperature on the curvature of space-time fabric*

First of all we recall our hypothesis resulting from the analysis of gravitational waves. The deformations of the space seem to be decoupled between the transverse deformations in the $(xy)$ plans and the distance z of the waves propagation. Thus $(ct\ and\ z)$ act at a same level(in the same direction) in a TT gauge in the expression of the deformations of gravitational waves. What leads T. Tenev and M.H Horstemeyer assumed the space consisting of thin sheets thick of Planck. I quote **[12]**.

"The thickness must be very small so that the fabric can behave as an essentially 3D object at ordinary lengthscales and be an appropriate analogy of 3D physical space. The thickness itself defines a microscopic lengthscale at which the behavior of the physical world would have to differ significantly from our ordinary experience. A value equal or comparable to Planck's length $l_p$ meets this criteria. However, the exact value of the thickness is not essential to the model as long as it is small but notvanishingly so".

We place ourselves at a distance from the cosmic web so as to consider on a given sheet of space a constant average temperature applied to said sheet (it is actually necessary to dissociate the quantum, local and global thermal effect associated with the scale factor, great thermal curvature of space) .which allows us to guard against the inevitable thermal variations at smaller scales associated with the cosmic web.

These two postulates in place we have:

A first effect associated with our space thermal expansion coefficient which implies a thermal expansion of the space in the plane $xy$ (see figure 4). The curvature of the space k being almost zero, it appears to us as a leaf that expands in part because of the average temperature of the cosmic web.

A second effect associated with our time thermal expansion coefficient involves this time a following thermal curvature linked this time between the temperature difference between two layers of Planck thickness space folowing the z direction.

The figure 4 below give an idea of the concepts developed in this paper.

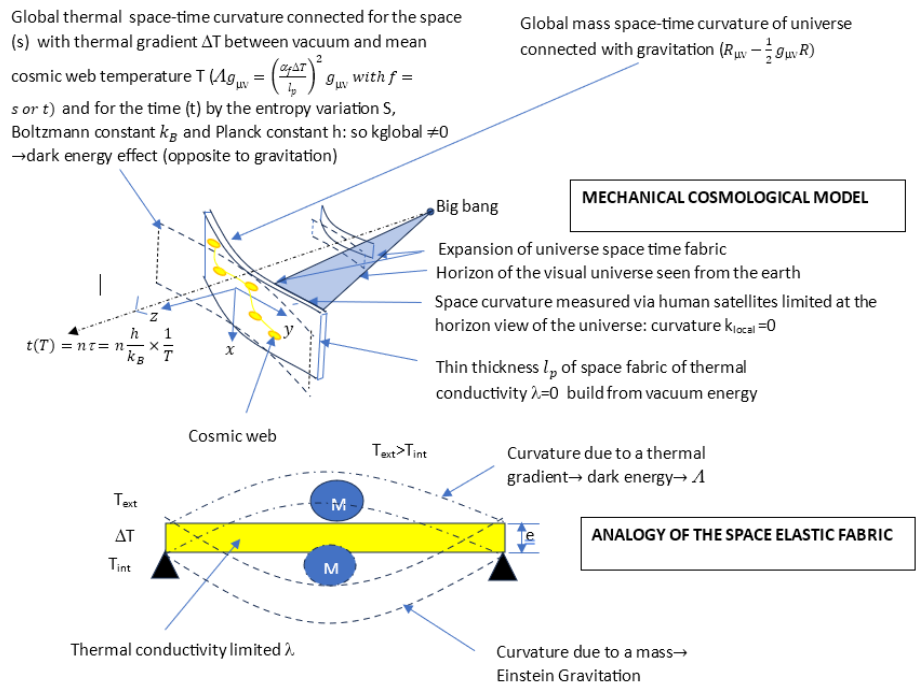

**Figure 4: Overview of the principles and concepts developed in this paper (thermal curvature of the space-time connected to Λ, time, entropy variation and temperature) -**

So in our model illustrated in Figure 4 in a way near to **[31]**:

1) Space-time expands as time progresses every moment since the Big bang,

2) The engine of time, the fact that each present moment is renewed indefinitely is correlated with entropy variation which can only increase over time: $t = \frac{S'}{\dot{S}}$,

3) Time function of temperature because linked to entropy variation itself is quantified by: $t_{(T)} = n\tau = n\frac{h}{k_B} \times \frac{1}{T}$,

4) The entropy S is linked to a temperature effect by the Bolzmann constant and by $\Omega$ different microstates by (S=$k_B \log \Omega$),

5) Space expands and via the interval $ds^2$ curves not only in the presence of mass (by general relativity) but also as a function of temperature or temperature delta (contribution of the analogy of general relativity with the theory of elasticity) which we propose to correlate with the cosmological constant $\Lambda$ as thermal curvature of quantum space sheets of thickness $l_p$,

6) By general relativity not only is space curved, but so is space-time, so seeing what precedes time ($\times c$) expands and curves by the effect of variation in entropy itself linked to an evolution of temperature **[27 ]**, **[29]**, **[31]**.

7) The curvature of space-time due to mass is opposed to the curvature of space time due to temperature namely the curvature contribution that is identified by the cosmological constant $\Lambda$.

**9. Presentation of the possible consequences of this model concerning the interpretation of dark energy**

The publications of **[19]** and **[20]** in particular and many others seem to indicate that the dark energy source of expansion of the universe is connected to the cosmological constant in particular when this one is positioned on the right of the equations (15) and (16). We are interested in this article in a cosmological constant placed on the left of the equation (17) that is to say in an additional curvature present in all space-time. This more mechanistic approach makes it possible to no longer have recourse to an energy of unknown origin and is closer to a functioning of the universe as a structure "charged" within it by the cosmic web and undergoing the thermal gradient of this one.

But this approach raises other questions (which thermal gradient considered? does space really have a sheets structure deforming relative to each other as the deformations associated with the polarizations $A^+$ and $A^\times$ that seem to suggest the gravitational waves? What is the real thickness of these sheets?how to integrate the anisotropy of space suggested by a Poisson's ratio of 1?The quantum field theory approach seems more coherent because it leads to a coeffcient of thermal expansion compatible with materials on Earth, nevertheless the difference with the astrophysical value of the cosmological constant raises a real fundamental question.Moreover the values of the Young's moduli that we recalled at the beginning of the article of on the one hand and the vacuum energy densities on the other hand being so outside the usual values of elastic materials on Earth that it is advisable to be particularly careful with respect to this "quasi-normal" value of the coefficient of expansion thermal that we propose in this study.

Another consequence of our study is that given that general relativity applies to space time, introducing a link between the cosmological constant and a temperature effect implies that temperature influences also time.

We therefore have by interpreting the expression (64), on the right a spatial part which only varies as a function of the temperature T and the coefficient of spatial thermal expansion $\alpha_s$ which nowadays is that of the vacuum (cold) in interaction with the temperature of the cosmic web (warm) which creates thermal gradient and curvature (so rather stable to day in mean), while the left part linked to time and to the coefficient of time thermal expansion $\alpha_t$ has only increased since the big bang. Thus the curvature term in (65) $-(\alpha_f T/l_p)^2$ represents the thermal curvature which is opposed at the gravitational curvature due to mass.

The part of the thermal curvature associated with time could have an apparent effect of accelerating the expansion of space. In (65) the global thermal space time curvature is so in our model in the opposite direction of mass curvature to play the game of the dark energy.

The temperature of the cosmic web also reflected an expansion of space in its $xy$ plane (see Figure 4) with (k apparent =0).

So, our approach is similar to **[31]** but using a mechanical formalism based on thermal expansion coefficient of the space time fabric.

Following our reflection, we could have a general curvature of space-time linked to temperature which is broken down into 2 parts. A first spatial part due to a thermal gradient applied to the thin sheets of space fabric which could be linked to the cosmological constant Λ (expansion of the universe and positive curvature k). A second resulting from a time/entropy/temperature effect. Both creating a total negative thermal curvature k opposite to the curvature linked to mass and therefore to gravitation.

10. How to verify this effect of the temperature on the time

Measuring the variation of time as a function of temperature on atomic clocks is a good way to test the ideas of this paper.

The idea is to put two strictly identical atomic clocks under the same height conditions in the earth's gravity field and to vary the temperature for one of them in order to measure the possible time lag with respect to the remaining one. at the initial temperature.

## 11. Conclusions

We explore in this study the analogy of space as an elastic medium by focusing on the mechanical parameters associated with any elastic four-dimensional fabric, its coefficients of thermal expansion $\alpha_s$ for space and $\alpha_t$ for the time. The state of the art is rather poor or even non-existent on their definitions and their intensities within the framework of an elastic model of space. Some research exists for the time dilatation function of temperature. By taking inspiration from models of space fabric in the form of thin sheets, placing ourselves within the framework of quantum field theory both on the value of the cosmological constant $\Lambda$ and on the thickness of these sheets of the order of the Planck's length, considering the recent measurements making it possible to establish the orders of magnitude of the space thermal gradient between the hot and cold zones of the universe and finally considering that these sheets bend under this thermal gradient by analogy with mechanics plates, we propose a formulation and a value of the expansion coefficients of the space fabric.

Since general relativity is built on spacetime, this thermal curvature approach implies that not only space but also time could vary with temperature. We therefore propose an expression for the temporal expansion coefficient $\alpha_t$ based on the ratio between Botzmann's constant and Planck's constant in connection with a variation in entropy of the universe and therefore its temperature. Since general relativity is itself based on special relativity, this results in the need to take into account variations in distance and time as a function of temperature. We therefore propose a way to modify the definition of the space-time invariant $ds^2$ by introducing both a space-time coefficient and a temporal expansion coefficient.

When we return to the simplistic analogy cited at the beginning of the stretched fabric supporting a heavy ball, it indeed appears logical that the fabric supporting this ball elongates more or less depending on its temperature in addition to spatial flexibility and mass/energy which distorts it.

Thus gravitation influences space and time but would also be linked to temperature. Temperature would influence space-time curvature at the quantum scale (possible nature of the time expansion coefficient) and at large scale (possible cosmological constant effect in link with the cosmic web). The answer in our elastic medium analogy would be of two types. A first on a lengthening and shortening of space time, a second on a thermal curvature of space time connected with the dark energy.

This study remains a initial pproach which somehow clears this path given the great variability of the intensity of the cosmological constant depending on whether one considers its value from quantum field theory or its value from cosmological measurements. Other modelings are certainly necessary to fine-tune the value of these thermal expansion coefficients of the space-time fabric and this study remains only a first approach.

Einstein said that God is not playing dice when he talks about quantum mechanics, but perhaps God is playing structural engineer with a cosmic structure driven by a thermal gradient manifesting as dark energy acting throughout the universe – at least that's the question this article poses.

**Acknowledgements**

Thank you finally to the late R Gregoire, a great mechanician, who through his teaching based on the research of "how it works" guided me in my reflection.

We warmly thank the reviewer for all their suggestions and corrections, in particular on the need to distribute this thermal effect both spatially and temporally.

**References**

[1] F.W. Dyson, A. S. Eddington and C. Davidson, *Phil. Trans. Of. The. R. Soc. Lond.* **220** 291 (1929)

[2] C.W. F. Everitt, *Phys. Rev. Let.* **106** (2011) 221101

[3] Collective and al, *Phys. Rev. Let.* **116** (2016) 061102

[4] Collective and al, *Phys. Rev. Let.* **119** (2017) 161101

[5] A. D. Sakharov, *Sov. Phys. Dokl.* **12** (1968) 1040

[6] J. L. Synge, *Math. sch. of. phy. sc.* **72** (1959) 82

[7] C. B. Rayner and R. Proc, *Soc. A. Math. Phys. And. Eng.* **272** (1963) 44

[8] R. Grot and A. Eringen, *Int. J. Eng. Sc.* **2** (1966) 1

[9] V. V. Vasiliev and L. V. Fedorov, *Mech. of. Sol.* **53** (2018) 256

[10] V.V. Vasilev and L.V. Fedorov, *Mech. of. Sol.* **56** (2021) 404

[11] J. D. Brown, *Class. And. Quan. Grav.* **38** (2021) 085017

[12] T. G. Tenev and M. F. Horstemeyer, *Int. J. Of. Mod. Phy.* D **27** (2018) 1850083

[13] P. A. Millette, *Elastodynamic of the space time continuum* (NewMexico, USA, STCED, American, Research, Press, NewYork, 2019)

[14]     M. Beau, *On the acceleration of the expansion of a cosmological medium* (Boston, US, Univ, Massachusetts, Boston, 2018)

[15]     K. McDonald, *What is the Stiffness of Spacetime* (Princeton, J, Henry, Lab, Princ, Univ, 2018)

[16]     D. Izabel, *Pram. J. Of. Phys.* **94** (2020) 119

[17]     A. C. Melissinos, *Upper limit on the Stiffness of space-time* (Rochester, Dept, Phys, Astro, Univ, Rochester, 2018)

[18]     S. R. Wang, *Estimation of spacetime stiffness based on LIGO observations* (Taiwan, ,Keelung, univ, 2003)

[19]     P. J. E. Peebles and B. Ratra, *Rev.of. Mod. Phys.* **75** (2003) 559

[20]     E. Santos, *Astr. Sp. Sc.* **332** (2011) 423

[21]     C. F. Gauss, *Disquisitiones generales circa superficies curvas* (Gottingen, Societate, Regiae, Oblatae, D, Gottingae, 1828)

[22]     H. A. Perko, *J. Phy. Conf. series.* **1956** (2021) 1

[23]     Collectif and al, *Astron. And. Astrop. Rev.* **571** (2014) 1

[24]     D. Izabel, *What is space time made of?* ( Paris, Edp, sciences, 2021)

[25]   S. Timoshenko and J.N. Goodier, *Theory of elasticity* (NewYork, McGraw, Hill, 1951)

[26]   Y. K. Chiang, R. Makiya, B. Ménard and E. Komatsu, *Astr. J.* **902** (2020) 56

[27]     U. Lucia, G. Grisolia, *Results. in. Physics.***16** (2020) 102977
[28]      H.Hadi, K.Atazadeh, F.Darabi, *Physics. Letters. B*. **834** (2022) 137471
[29]     A. Chatterjee,G. Iannacchione, *arXiv:2007.09398v1* **1** *(2020) 1*

[30]     T.Tenev, *thesis An elastic constitutive model of spacetime and its applications* (Mississippi State, Mississippi, univ 2018)

[31]     C. Hauret, P. Magain, and J. Biernaux,*MDPI. Entropy.* **19** (2017) 357
[32]     CHAS. A. Eganc, H. Lineweaver, *Astr. J*. **710** (2010) 1825–1834